\documentclass[conference]{IEEEtran}
\IEEEoverridecommandlockouts
\usepackage{cite}
\usepackage{amsmath,amssymb,amsfonts}
\usepackage{algorithmic}
\usepackage{algorithm}

\usepackage{graphicx}
\usepackage{textcomp}
\usepackage{soul,color,xcolor}
\def\BibTeX{{\rm B\kern-.05em{\sc i\kern-.025em b}\kern-.08em
		T\kern-.1667em\lower.7ex\hbox{E}\kern-.125emX}}

\usepackage{booktabs}

\usepackage{tabularx}
\usepackage{array}
\usepackage{graphicx}
\usepackage{float}
\usepackage{subfig}

\makeatletter
\newcommand{\linebreakand}{%
\end{@IEEEauthorhalign}
\hfill\mbox{}\par
\mbox{}\hfill\begin{@IEEEauthorhalign}
}
\makeatother

\begin{document}
\title{Dispatch-Embedded Long-Term Tail Risk Assessment and Mitigation via CVaR for Renewable Power Systems
\\
\thanks{
This work was supported by the Joint Research Fund in Smart Grid (No.U1966601) under cooperative agreement between the National Natural Science Foundation of China (NSFC).
}
}
\author{\IEEEauthorblockN{Kai Kang, Feng Liu*}
\IEEEauthorblockA{\textit{Department of Electrical Engineering, Tsinghua University,	Beijing, China}\\ 
kk21@tsinghua.org.cn }
}
\maketitle
\begin{abstract}
Renewable energy (RE) generation exhibits pronounced seasonality and variability, and neglecting these features can lead to significant underestimation of long-term power system risks in power supply. While long-term dispatch strategies are essential for evaluating and mitigating tail risks, they are often excluded from existing models due to their complexity. This paper proposes a long-term tail risk assessment and mitigation framework for renewable power systems, explicitly embedding dispatch strategies. 
A representative scenario generation method is designed, combining multi-timescale Copula modeling to capture RE's long-range variability and correlation. Building on these scenarios, an evolution-based risk assessment model is established, where Conditional Value-at-Risk (CVaR) is employed as a robust metric to quantify tail risks. Finally, a controlled evolution-based risk mitigation scheme is introduced to refine long-term dispatch strategies for mitigating tail risks. Case studies on a modified IEEE-39 bus system incorporating real-world data substantiate the efficacy of the proposed method.

\end{abstract}


\section{Introduction}
 
The environmental and economic benefits of renewable energy (RE) drive its growing integration into modern power systems. However, RE exhibits seasonality and strong dependence on weather conditions. For instance, due to variations in solar radiation, photovoltaic (PV) generation in Germany during winter typically amounts to only 20–40\% of that in summer, while in the United Kingdom, wind power shows a strong seasonal profile, with average capacity factors approaching 45\% in winter but declining to below 20\% in summer \cite{staffell2023global,kang2025extreme}.  Neglecting these long-term characteristics can result in a substantial underestimation of power supply tail risks \cite{xu2024resilience}, underscoring the urgent need for advanced long-term risk assessment and mitigation methods.

Accurately characterizing RE uncertainty is essential for risk assessment. On short timescales, uncertainty can be modeled using forecasting techniques such as ARIMA models, machine learning, or statistical resampling \cite{shumway2017arima,shayan2024innovative,james2023resampling}, and expressed by uncertainty sets, probabilistic scenarios, or fuzzy distributions. 
In contrast, long-term scenario generation must capture seasonal patterns and the possibility of extreme events, which are difficult to predict. Traditional clustering-based methods \cite{li2020clustering} provide useful insights but struggle to represent joint extreme events. Monte Carlo simulation \cite{urbanucci2018optimal} can generate numerous samples but struggles with efficiently capturing tail risks. Copula-based methods \cite{fan2021uncertainty} offer improved modeling of joint extreme events, yet face computational challenges in high dimensions. 

Currently, most risk assessment models focus on short-term horizons, addressing uncertainties in renewable generation, load forecasting errors, and extreme weather events \cite{liang2019risk,li2025risk,sun2025resilience}. However, these methods often fail to capture persistent extreme conditions, such as prolonged low-wind periods or cloudy seasons, that significantly influence system risk but remain invisible in short-term analyses. 

Traditional long-term risk assessments rely on simplified methods such as power duration curves \cite{pineda2018chronological}, which cannot adequately address the increasing RE variability. More recent advances include optimization-based approaches, such as stochastic optimization (SO) \cite{lara2021multi}, robust optimization (RO) \cite{kumar2021novel}, and distributionally robust optimization (DRO) \cite{yang2024distributionally}, typically combined with risk metrics like loss-of-load probability (LOLP) \cite{rashidaee2018linear}, expected unserved energy (EUE) \cite{milligan2019capacity}, and other reserve adequacy indices \cite{kang2025extreme}. These methods often assume idealized dispatch, overlooking realistic long-term operational strategies that crucially affect system risks, especially tail risks, under uncertainties \cite{kang2024enforcing}. 
Consequently, they may misestimate system risks and provide limited insight for risk mitigation.

To address these gaps, this paper proposes a comprehensive framework for dispatch-embedded long-term tail risk assessment and mitigation. 
First, a representative scenario generation approach is developed, integrating multi-timescale Copula modeling to capture long-range dependencies and variability in RE. Subsequently, a long-term risk assessment model is formulated, embedding dispatch strategies and applying Conditional Value-at-Risk (CVaR) to quantify tail risks. Finally, a controlled evolution-based risk mitigation method is proposed to refine the long-term dispatch strategies.

\section{Renewable Power System Model}
\subsection{Outline}
\begin{figure}[htb]
	\centering
	\includegraphics[width=9cm]{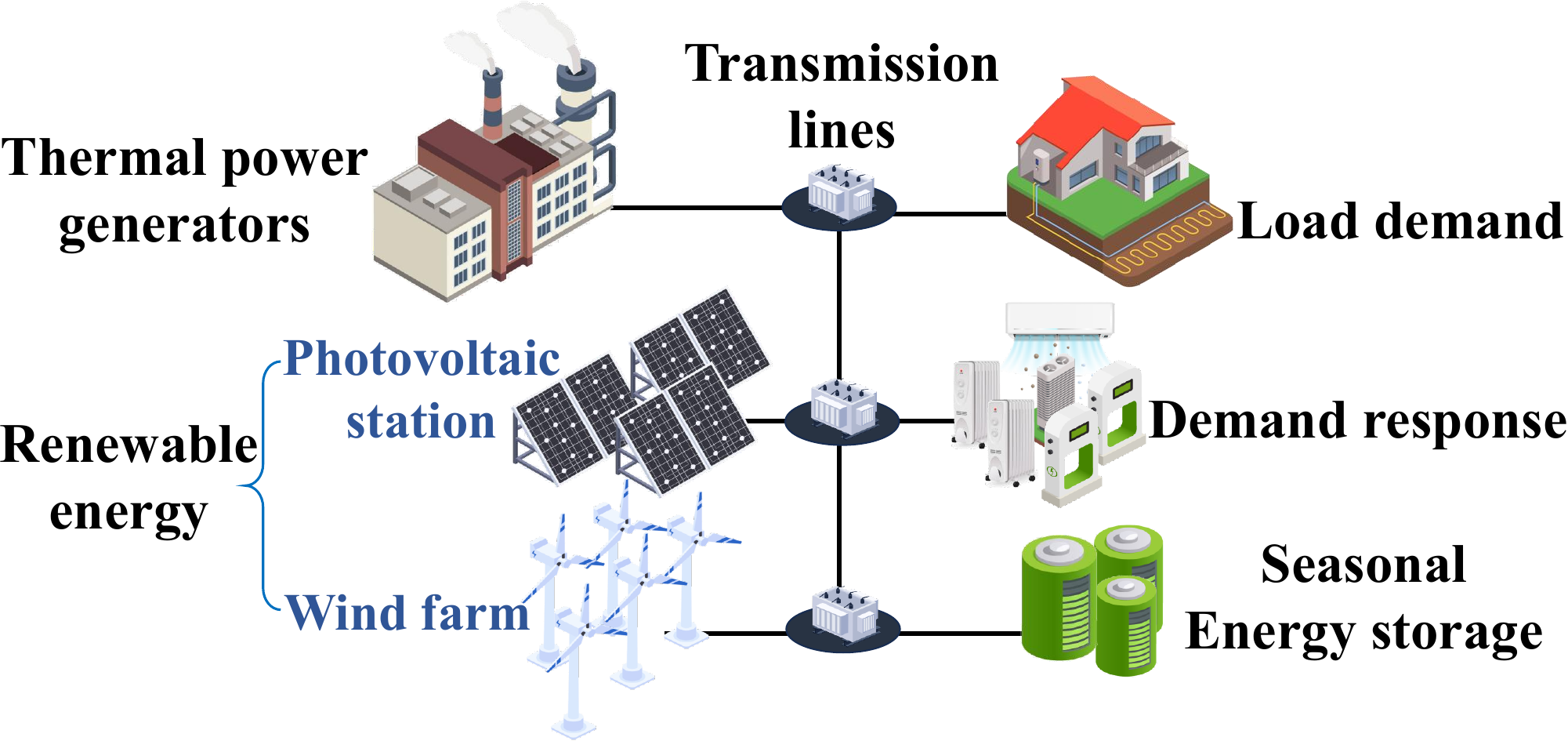}
	\caption{Renewable power system diagram.}\vspace{-0.4cm}
	\label{Fig:System_diagram}
\end{figure}

The renewable power system consists of thermal power generators (TPGs), renewable energy (RE), seasonal energy storage (SES), demand response (DR),  load demand (LD), and transmission lines (TLs), as shown in Fig. \ref{Fig:System_diagram}. Each type of equipment is represented by a set $\Omega_\bullet$, where $\bullet$ corresponds to TPG, RE, SES, DR, LD, or TL, respectively.
The time period horizon is denoted by set $\Omega_\mathrm{T}$, with index $t$ and number $\left|\Omega_\mathrm{T}\right|$. 
\subsection{System Constraints}

\subsubsection{TPG} Each TPG $g\in\Omega_\mathrm{TPG}$ is subject to:
\begin{subequations}
	\label{eq:TPG_cons}
	\setlength\abovedisplayskip{1pt}
	\setlength\belowdisplayskip{2pt}
	\begin{gather}
		\label{eq:TPG_1}
		({x_{g,t}^\mathrm{TPG}} - {x_{g,t + 1}^\mathrm{TPG}})(X_{g,t}^\mathrm{ON} - T_g^\mathrm{ON}) \ge 0
		\\
		\label{eq:TPG_2}
		({x_{g,t}^\mathrm{TPG}} - {x_{g,t + 1}^\mathrm{TPG}})(X_{g,t}^\mathrm{OFF} - T_g^\mathrm{OFF}) \ge 0
		\\
		\label{eq:TPG_3}
		{y_{g,t + 1}^\mathrm{TPG}} - {z_{g,t + 1}^\mathrm{TPG}} = {x_{g,t + 1}^\mathrm{TPG}} - {x_{g,t}^\mathrm{TPG}}
		\\
		\label{eq:TPG_4}
		{x_{g,t}^\mathrm{TPG}},{y_{g,t}^\mathrm{TPG}},{z_{g,t}^\mathrm{TPG}} \in \left\{0,1\right\}
		\\
		\label{eq:TPG_5}
		\underline{P}_g^\mathrm{TPG}{x_{g,t}^\mathrm{TPG}} \le {p_{g,t}^\mathrm{TPG}} \le \overline{P}_g^\mathrm{TPG}{x_{g,t}^\mathrm{TPG}}
		\\
		\label{eq:TPG_6}
		{p_{g,t}^\mathrm{TPG}} - {p_{g,t + 1}^\mathrm{TPG}} \le {R_g^\mathrm{D}}{x_{g,t + 1}^\mathrm{TPG}} + \overline{P}_g^\mathrm{TPG}(1 - {x_{g,t + 1}^\mathrm{TPG}})
		\\
		\label{eq:TPG_7}
		{p_{g,t + 1}^\mathrm{TPG}} - {p_{g,t}^\mathrm{TPG}} \le {R_g^\mathrm{U}}{x_{g,t}^\mathrm{TPG}} + \overline{P}_g^\mathrm{TPG}(1 - {x_{g,t}^\mathrm{TPG}})
	\end{gather}
\end{subequations}
where binary variables $x_{g,t}^\mathrm{TPG}$, $y_{g,t}^\mathrm{TPG}$, and $z_{g,t}^\mathrm{TPG}$ indicate commitment, startup, and shutdown states. Variable $p_{g,t}^\mathrm{TPG}$ is the dispatched power of TPG $g$. 

Constraints \eqref{eq:TPG_1}-\eqref{eq:TPG_2} enforce minimum on-time $T_g^\mathrm{ON}$ and off-time $T_g^\mathrm{OFF}$ durations, based on the accumulated on and off durations $X_{g,t}^\mathrm{ON}$ and $X_{g,t}^\mathrm{OFF}$. The relationship among $x_{g,t}^\mathrm{TPG}$, $y_{g,t}^\mathrm{TPG}$, and $z_{g,t}^\mathrm{TPG}$ is given by \eqref{eq:TPG_3}. TPG generation limits $[\underline{P}_g^\mathrm{TPG},\overline{P}_g^\mathrm{TPG}]$ are enforced by \eqref{eq:TPG_5}. Ramp-down/up limits $R_g^\mathrm{D}$ and $R_g^\mathrm{U}$ are captured by \eqref{eq:TPG_6}-\eqref{eq:TPG_7}. 

\subsubsection{RE}
Each RE $r\in\Omega_\mathrm{RE}$ satisfies:
\begin{equation}
	\label{eq:RG_1}
		\setlength\abovedisplayskip{1pt}
	\setlength\belowdisplayskip{2pt}
	0 \le p_{r,t}^\mathrm{RE} \le \overline{P}_{r,t}^\mathrm{RE}
\end{equation}
where variable ${p_{r,t}^\mathrm{RE}}$ is the utilized RE power, and $\overline{P}_{r,t}^\mathrm{RE}$ is the available RE generation.

\subsubsection{SES}
Each SES $e\in\Omega_\mathrm{SES}$ is governed by:
\begin{subequations}
	\label{eq:ES_cons}
	\begin{gather}
		\label{eq:SES_1}
		0 \le p_{\mathrm{CH},e,t}^\mathrm{SES} \le {a_{e,t}^\mathrm{SES}} \overline{P}_{e}^{\mathrm{SES}}
		\\
		\label{eq:SES_2}
		0 \le  p_{\mathrm{DC},e,t}^\mathrm{SES} \le (1-{a_{e,t}^\mathrm{SES}})\overline{P}_{e}^{\mathrm{SES}}
		\\
		\label{eq:SES_3}
		{a_{e,t}^\mathrm{SES}} \in \left\{0,1\right\}
		\\
		\label{eq:SES_4}
		w_e^\mathrm{D}\overline{E}_{e}^{\mathrm{SES}} \le E_{e,t}^\mathrm{SES} \le  w_e^\mathrm{U}\overline{E}_{e}^{\mathrm{SES}}
		\\
		\label{eq:SES_5}
		{E_{e,t}^\mathrm{SES}} = {E_{e,t-1}^\mathrm{SES}} + \left(\eta_e^\mathrm{CH}p_{\mathrm{CH},e,t}^\mathrm{SES} - p_{\mathrm{DC},e,t}^\mathrm{SES}/\eta_e^\mathrm{DC}\right)\Delta t
		\\
		\label{eq:SES_6}
		{E_{e,0}^\mathrm{SES}} = E_{\mathrm{INT},e}^\mathrm{SES}
		\\
		\label{eq:SES_7}
		{E_{e,\left|\Omega_{\mathrm{T}} \right|}^\mathrm{SES}} \ge {E_{e,0}^\mathrm{SES}}
	\end{gather}
\end{subequations}
where variables $p_{\mathrm{CH},e,t}^\mathrm{SES}$ and  $p_{\mathrm{DC},e,t}^\mathrm{SES}$ are charge and discharge power, and $\overline{P}_{e}^{\mathrm{SES}}$ is the power capacity. Binary variable $a_{e,t}^\mathrm{SES}$ prevents simultaneous charge and discharge.  $E_{e,t}^\mathrm{SES}$ is the state of charge (SoC); $w_e^\mathrm{D}$ and $w_e^\mathrm{U}$ are upper/lower bound coefficients for energy capacity $\overline{E}_{e}^{\mathrm{SES}}$. SoC dynamics are defined by \eqref{eq:SES_5}, with $\eta_e^\mathrm{CH}$ and $\eta_e^\mathrm{DC}$ as charge/discharge efficiencies, and $\Delta t$ as a time period duration. 
$E_{\mathrm{INT},e}^\mathrm{SES}$ is the initial SoC, and the final SoC must not fall below $E_{\mathrm{INT},e}^\mathrm{SES}$, as per 
\eqref{eq:SES_7}. 

%

\subsubsection{DR} Each DR $m\in\Omega_\mathrm{DR}$ is subject to:
\begin{equation}
	\label{eq:DR_cons}
	\setlength\abovedisplayskip{1pt}
	\setlength\belowdisplayskip{2pt}
	0 \le p_{m,t}^\mathrm{DR} \le \overline{P}_{m}^\mathrm{DR}
\end{equation}
where variable $p_{m,t}^\mathrm{DR}$ is the utilized DR power and $\overline{P}_{m}^\mathrm{DR}$ is the DR capacity.

\subsubsection{Power Supply}
The system power supply is:
\begin{subequations}
	\label{eq:PowerBalance_cons}
	\setlength\abovedisplayskip{1pt}
	\setlength\belowdisplayskip{2pt}
	\begin{gather}
		\label{eq:PowerBalance_1}
		\begin{array}{c}
			\sum\limits_{g \in {\Omega _\mathrm{TPG}}} {p_{g,t}^\mathrm{TPG}}
			+ \sum\limits_{r \in {\Omega _\mathrm{RE}}} {p_{r,t}^\mathrm{RE}}
			+\sum\limits_{m \in {\Omega _\mathrm{DR}}} p_{m,t}^\mathrm{DR} 
			\\ 	
			+ \sum\limits_{e \in {\Omega _\mathrm{SES}}} {\left( p_{\mathrm{DC},e,t}^\mathrm{SES} -  p_{\mathrm{CH},e,t}^\mathrm{SES}\right)}		
			 + p_{t}^\mathrm{B+}
			\ge \sum\limits_{d \in {\Omega _\mathrm{LD}}} {\overline{P}_{d,t}^\mathrm{LD}}
		\end{array}
		\\
		\label{eq:PowerBalance_2}
		p_{t}^\mathrm{B+} \ge 0
	\end{gather}
\end{subequations}
where $\overline{P}_{d,t}^\mathrm{LD}$ is the demand of LD $d\in\Omega_\mathrm{LD}$. Variable $p_{t}^\mathrm{B+}$ is the non-negative emergency power support, representing system's power shortages.

\subsubsection{TL}
The power flow for each TL $l\in\Omega_\mathrm{TL}$ is limited by:
\begin{equation}
	\label{eq:TL_cons}
	\setlength\abovedisplayskip{1pt}
	\setlength\belowdisplayskip{2pt}
	\begin{array}{c}
		\left| 
		{\begin{array}{*{3}{c}}
				{
				\sum\limits_{g \in {\Omega _\mathrm{TPG}}} {{\rm{SF}}_{l,g}^\mathrm{TPG} {p_{g,t}^\mathrm{TPG}}} 
				+\sum\limits_{r \in {\Omega _\mathrm{RE}}} {{\rm{SF}}_{l,r}^\mathrm{RE} p_{r,t}^\mathrm{RE}}
				}				
				\\	
				+ 
				\sum\limits_{e \in \Omega _\mathrm{SES}} {\rm{SF}}_{l,e}^\mathrm{SES}\left(p_{\mathrm{DC},e,t}^\mathrm{SES} -  p_{\mathrm{CH},e,t}^\mathrm{SES} \right) 
				\\
				{
				 + \sum\limits_{m \in {\Omega _\mathrm{DR}}} {{\rm{SF}}_{l,m}^\mathrm{DR} p_{m,t}^\mathrm{DR}}
				- \sum\limits_{d \in {\Omega_\mathrm {LD}}} {\rm{SF}}_{l,d}^\mathrm{LD} \overline{P}_{d,t}^\mathrm{LD} 
				}
		\end{array}} 
		\right|
		\le {\overline{\mathrm{PL}} _l}
	\end{array}
\end{equation} 
where parameters ${\rm{SF}}_{l,i}^\bullet$ represents the shift factor of TL $l$ for equipment $i \in \Omega_\bullet$, and ${\overline{\mathrm{PL}} _l}$ is the TL capacity. 

\subsection{Objective Function}
In the renewable power system, the cost for period $t$ consists of the TPG cost $L_t^\mathrm{TPG}$, ES cost $L_t^\mathrm{SES}$ \eqref{eq:L_SES}, DR cost $L_t^\mathrm{DR}$ \eqref{eq:L_DR}, and emergency supply cost $L_t^\mathrm{EME}$ \eqref{eq:L_EME}:
\begin{subequations}
	\label{eq:L_obj}
	\setlength\abovedisplayskip{1pt}
	\setlength\belowdisplayskip{2pt}
	\begin{gather}
		\label{eq:L_TPG}
		L_t^\mathrm{TPG} = \sum\nolimits_{g}\left(
		c_{g}^\mathrm{SU} y_{g,t}^\mathrm{TPG}+c_{g}^\mathrm{SD} z_{g,t}^\mathrm{TPG}
		+c_{g}^\mathrm{TPG} p_{g,t}^\mathrm{TPG}\Delta t
		\right)
		\\
		\label{eq:L_SES}
		L_t^\mathrm{SES} = \sum\nolimits_{e} c_e^{{\rm{SES}}}\left(p_{\mathrm{CH},e,t}^\mathrm{SES} + p_{\mathrm{DC},e,t}^\mathrm{SES}\right)\Delta t
		\\
		\label{eq:L_DR}
		L_t^\mathrm{DR} = \sum\nolimits_{m}c_m^{\mathrm{DR}} p_{m,t}^\mathrm{DR} \Delta t
		\\
		\label{eq:L_EME}
		L_t^\mathrm{EME} = {c^\mathrm{EME}} p_{t}^\mathrm{B+}\Delta t
	\end{gather}
\end{subequations}
where $c_{g}^\mathrm{SU}$, $c_{g}^\mathrm{SD}$, $c_{g}^\mathrm{TPG}$, $c_{e}^{{\rm{SES}}}$, and $c_{m}^{{\rm{DR}}}$ are respective prices for each component, while $c^\mathrm{EME}$ is the emergency supply price.

\section{Long-Term Representative Scenario Generation}
We denote the historical scenario set as $\Omega_\mathrm{HST}$ with quantity $N_\mathrm{HST}$, and each scenario $h$ is defined as:
\begin{equation}
	\notag
	\setlength\abovedisplayskip{1pt}
	\setlength\belowdisplayskip{2pt}
		\overline P_h^{{\rm{HST}}} = \left\{ \left\{\overline P_{r,t,h}^{\rm{RE}}\right\} _{r \in \Omega_\mathrm{RE} },  \, \, \left\{ \overline P_{d,t,h}^{\rm{LD}} \right\} _{d \in \Omega_\mathrm{LD}}   \right\}_{t = 1}^{\left|\Omega_\mathrm{T}\right|},
\end{equation}
where $\overline P_{r,t,h}^{\rm{RE}}$ and $ \overline P_{d,t,h}^{\rm{LD}}$ represent the renewable generation and load demand of scenario $h$ at period $t$, respectively.

To efficiently capture both long-term trends and short-term variability, we propose a multi-timescale Copula-based scenario generation approach, summarized in Algorithm \ref{alg:scenario_generation}. While high-dimensional Copula modeling is computationally challenging, it mitigates dimensionality by decomposing scenarios into a long-term horizon (modeled via Copula on aggregated data) and a short-term horizon (restored using sequential sampling), ensuring generated scenarios reflect both large-scale correlations and fine-grained temporal variability.
\begin{algorithm}[htb]
     \small 
	\caption{Multi-timescale Copula-based long-term representative scenario generation.} 
	\label{alg:scenario_generation} 
	\textbf{Input:} 
	Historical scenario set $\Omega_\mathrm{HST}$; desired number of scenarios $N_\mathrm{REP}$. Aggregation length $A_\mathrm{G}$; number of aggregation blocks $K_\mathrm{G}$, where $A_\mathrm{G}\cdot K_\mathrm{G} = \left|\Omega_\mathrm{T}\right|$. 
	\\
	\makebox[\linewidth]{\textbf{Long-term horizon: Aggregate scenario generation}}
	\\
	\textbf{Step 1:} 
	For each $h \in \Omega_\mathrm{HST}$, aggregate power over each block $k$:
	\\ 
	$\overline P_{r,k,h}^{\rm{RE,AG}} = \sum_{t=(k-1) A_\mathrm{G}+1}^{k A_\mathrm{G}}\overline P_{r,t,h}^{\rm{RE}} $ ($r \in \Omega_\mathrm{RE}$),
	\\
	$P_{d,k,h}^{\rm{LD,AG}} = \sum_{t=(k-1) A_\mathrm{G}+1}^{k A_\mathrm{G}}\overline P_{d,t,h}^{\rm{LD}} $ ($d \in \Omega_\mathrm{LD}$).
	\\
	Hence, each scenario becomes:
	\\
	$\overline P_h^{\mathrm{HST,AG}} = \left\{ \left\{\overline P_{r,k,h}^{\rm{RE,AG}}\right\} _{r \in \Omega_\mathrm{RE} },  \left\{\overline  P_{d,k,h}^{\rm{LD,AG}}\right\} _{d \in \Omega_\mathrm{LD}}   \right\}_{k = 1}^{K}$.
	 \\
	\textbf{Step 2:} 
	For each $r$, $d$, and $k$: fit the marginal distributions of $\left\{\overline P_{r,k,h}^{\rm{RE,AG}}\right\} _{h \in \Omega_\mathrm{HST} }$ and $\left\{ P_{d,k,h}^{\rm{LD,AG}}\right\} _{h \in \Omega_\mathrm{HST}} $. 
	\\
	\textbf{Step 3:} Fit a Copula to model dependencies among all aggregated variables.
	\\
	\textbf{Step 4:} Generate $N_\mathrm{REP}$ aggregated scenarios, and the $s$-th scenario:\\
	$\overline P_s^{\mathrm{REP,AG}} = \left\{ \left\{\overline P_{r,k,s}^{\rm{RE,AG}}\right\} _{r \in \Omega_\mathrm{RE} },  \left\{\overline P_{d,k,s}^{\rm{LD,AG}}\right\} _{d \in \Omega_\mathrm{LD}}   \right\}_{k = 1}^{K}$.
	\\
	\makebox[\linewidth]{\textbf{Short-term horizon: Detailed scenario supplement}}
	\textbf{Step 5:} For each $s$ and block $k$: \\Initialize $\overline P_{r, (k-1) A_\mathrm{G}+1 ,s}^{\rm{RE}}$ and $\overline P_{d,(k-1) A_\mathrm{G}+1,s}^{\rm{LD}}$ from historical marginal distributions. \\Fit a 2-D Copula to sequentially generate $\overline P_{r, t,s}^{\rm{RE}}$ and $\overline P_{d,t,s}^{\rm{LD}}$ for $t = (k-1) A_\mathrm{G}+2,..., k A_\mathrm{G}$.
	\\
	\textbf{Step 6:} Scale the $\left\{\overline P_{r, t,s}^{\rm{RE}}\right\}_{t=(k-1) A_\mathrm{G}+1}^{k A_\mathrm{G}}$ and $\left\{\overline P_{d,t,s}^{\rm{LD}}\right\}_{t=(k-1) A_\mathrm{G}+1}^{k A_\mathrm{G}}$ within block $k$ to match the aggregated Copula scenario: 
	\\ $\overline P_{r,t,s}^{\rm RE} \gets 
	\frac{	\overline P_{r,t,s}^{\rm RE} \overline P_{r,k,h}^{\rm{RE,AG}}}{\sum\nolimits_{t=(k-1) A_\mathrm{G}+1}^{k A_\mathrm{G}} \overline P_{r,t,s}^{\rm RE}}$, 
	$\overline P_{d,t,s}^{\rm{LD}} \gets 
	\frac{	\overline P_{d,t,s}^{\rm{LD}} \overline P_{d,k,s}^{\rm{LD,AG}}}{\sum\nolimits_{t=(k-1) A_\mathrm{G}+1}^{k A_\mathrm{G}} \overline P_{d,t,s}^{\rm{LD}}}$.
	\\
	\textbf{Step 7:} Concatenate $\overline P_{r,t,s}^{\rm RE}$ and $\overline P_{d,t,s}^{\rm{LD}}$ for all periods $t$ to form full scenario: $\overline P_s^{\rm REP}=\left\{ \left\{\overline P_{r,t,s}^{\rm{RE}}\right\} _{r \in \Omega_\mathrm{RE} },  \,  \left\{ \overline P_{d,t,s}^{\rm{LD}} \right\} _{d \in \Omega_\mathrm{LD}}   \right\}_{t = 1}^{\left|\Omega_\mathrm{T}\right|}$.
	\\
	\textbf{Output:} 
	Representative scenario set $\Omega_\mathrm{REP} = \left\{\overline P_s^{{\rm{REP}}}\right\}_{s\in [1,N_\mathrm{REP}]}$.
\end{algorithm}

\section{CVAR-Based Long-Term Risk Assessment Considering Dispatch Strategies}
Dispatch strategies are crucial for mitigating system risks, especially over long-term horizons where decisions span multiple time scales and involve long-term operation constraints. However, most existing long-term risk assessment methods neglect the influence of realistic dispatch strategies.

\begin{figure}[htb]
	\centering
	\includegraphics[width=9cm]{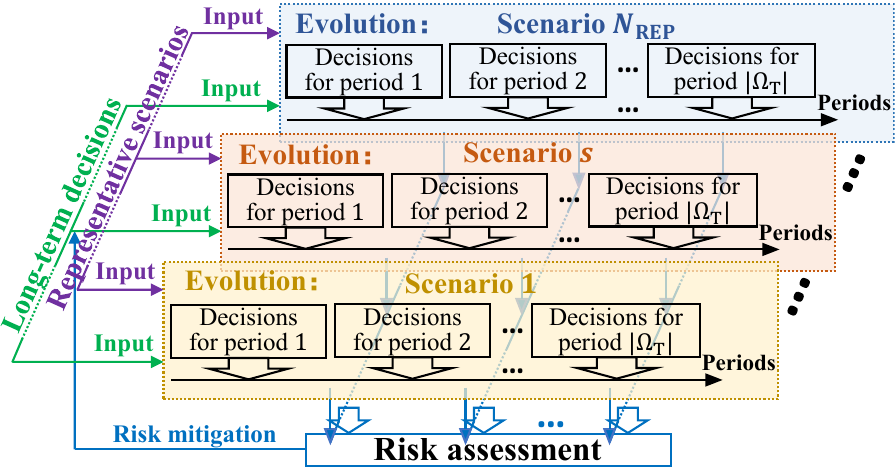}
	\caption{The evolution-based risk assessment model.}\vspace{-0.5cm}
	\label{Fig:assessment_framework}
\end{figure}

We propose an evolution-based risk assessment model, illustrated in Fig. \ref{Fig:assessment_framework}. It leverages the representative scenario set $\Omega_\mathrm{REP}$ and dispatch strategies to perform parallel time-sequential evolution and risk assessment.

For the renewable power system in Fig. \ref{Fig:System_diagram}, long-term decisions include determining the SoC sequence of SES, denoted as $\left\{\hat E_{e,t}^\mathrm{SES}\right\}_{t = 1}^{\left|\Omega_\mathrm{T}\right|}$, which can be easily obtained via two-stage dispatch methods \cite{wakui2022shrinking,kang2025securing}. Subsequently, short-term dispatch decisions are evolved within a rolling time window $\delta^\mathrm{R}$ for each representative scenario $s$. The short-term evolution at period $\tau$ is formulated as:
\begin{subequations}
	\label{eq:Evolution}
	\setlength\abovedisplayskip{1pt}
	\setlength\belowdisplayskip{2pt}
	\begin{gather}
		\label{eq:Evolution_obj}
		\min L_{s,\tau}^{\mathrm{E}} =  \sum\limits_{t =\tau}^{\tau+\delta^\mathrm{R}}\left\{
		\begin{array}{l}
			 L_t^\mathrm{TPG}+L_t^\mathrm{SES}+L_t^\mathrm{DR}+L_t^\mathrm{EME}
			  \\ + \omega^\mathrm{RT} \left[
			 E_{e,t}^\mathrm{SES} - \hat E_{e,t}^\mathrm{SES}
			\right]^2 
		\end{array}
		\right\}
		\\	
		\notag
		\text{ s.t. for $\forall t \in \left[\tau, \tau+\delta^\mathrm{R} \right]$: }
		\\
		\label{eq:Evolution_cons_1}
		\eqref{eq:TPG_cons}-\eqref{eq:TL_cons},\,
		\overline{P}_{r,t}^\mathrm{RE} = \overline P_{r,t,s}^{{\rm{RE}}}, \,
		\overline{P}_{d,t}^\mathrm{LD}= \overline P_{d,t,s}^{\mathrm{LD}}
	\end{gather}
\end{subequations}
where solutions of $L_{\tau}^\mathrm{EME}$ and $E_{e,\tau}^\mathrm{SES}$ are denoted as $\tilde{L}_{\tau,s}^\mathrm{EME}$ and $\tilde{E}_{e,\tau,s}^\mathrm{SES}$, and subscript $s$ is the scenario index. $\omega^\mathrm{RT}$ is a weighting coefficient penalizing deviations of $E_{e,t}^\mathrm{SES}$ from the long-term SoC sequence.

After the evolution using \eqref{eq:Evolution} for all scenarios $s$, the solutions $\left\{\tilde{L}_{t,s}^\mathrm{EME}\right\}_{t = 1}^{\left|\Omega_\mathrm{T}\right|}$ and $\left\{\tilde{E}_{e,t,s}^\mathrm{SES}\right\}_{t = 1}^{\left|\Omega_\mathrm{T}\right|}$ are obtained. For each period $t$, the $\alpha$\%-CVaR of emergency supply cost is computed as:
\begin{equation}
	\label{eq:CVaR}
	\setlength\abovedisplayskip{1pt}
	\setlength\belowdisplayskip{2pt}
	\begin{aligned}
		L_{\alpha\%,t}^\mathrm{CVaR} = L_{\alpha\%,t}^\mathrm{VaR} 
		+ \frac{\sum\nolimits_{s\in \Omega_\mathrm{REP}}\max(\tilde{L}_{t,s}^\mathrm{EME}-L_{\alpha\%,t}^\mathrm{VaR},0)}{\lceil (1-\alpha\%)\cdot N_\mathrm{REP}\rceil}
	\end{aligned}
\end{equation}
where $L_{\alpha\%,t}^\mathrm{VaR}$ is the $\alpha\%$ quantile of the sorted set ${\tilde{L}{t,s}^\mathrm{EME}}$. Thus, the $\left\{ L_{\alpha\%,t}^\mathrm{CVaR} \right\}_{t = 1}^{\left|\Omega_\mathrm{T}\right|}$ reflects the system's long-term risk levels. Comparing this CVaR sequence with the predefined risk threshold $T_{\alpha\%,t}^\mathrm{CVaR}$ allows identification of high-risk periods requiring mitigation. Long-term dispatch strategies can then be adjusted accordingly to reduce system tail risk. 

\section{Controlled Evolution-Based Risk Mitigation}
\label{Section:Risk_Mitigation}
When the CVaR at any period $t \in \Omega_\mathrm{T}$ exceeds the predefined threshold, i.e., $L_{\alpha\%,t}^\mathrm{CVaR} > T_{\alpha\%,t}^\mathrm{CVaR}$, the system is considered at elevated tail risk. To address this, we propose a controlled evolution-based risk mitigation method to refine the long-term dispatch strategy $\left\{\hat E_{e,t}^\mathrm{SES}\right\}_{t = 1}^{\left|\Omega_\mathrm{T}\right|}$.

In the short-term evolution \eqref{eq:Evolution}, the subgradient of $L_{s,\tau}^{\mathrm{E}}$ with respect to $\hat E_{e,\tau}^\mathrm{SES}$ is denoted as $G_{e,s}$, which is given by:  
\begin{equation}
	\label{eq:Subgradient_CE}	
	\setlength\abovedisplayskip{1pt}
	\setlength\belowdisplayskip{2pt}
	G_{e,s} = -2\omega^\mathrm{RT} \left[\tilde{E}_{e,t}^\mathrm{SES} - \hat E_{e,t}^\mathrm{SES}\right]
\end{equation}

This subgradient provides a descent direction for adjusting the reference SoC to mitigate risk. Let $\Omega_{\mathrm{REP},t}^{\mathrm{CVaR}}$ denote the subset of scenarios where the emergency supply cost exceeds the VaR, i.e., $\tilde{L}_{t,s}^\mathrm{EME} \ge L_{\alpha\%,t}^\mathrm{VaR}$.  Then, the condition for descending $\tilde{L}_{t,s}^\mathrm{EME}$ can be expressed as:
\begin{equation}
	\label{eq:Correct_region}
	\setlength\abovedisplayskip{1pt}
	\setlength\belowdisplayskip{2pt}
	\tilde{L}_{t,s}^\mathrm{EME} - L_{t}^\mathrm{EME} \le \sum\nolimits_{e \in \Omega _\mathrm{SES}} G_{e,s}\left[\hat E_{e,t}^\mathrm{SES}-E_{e,t}^\mathrm{SES}\right]
\end{equation}

Averaging \eqref{eq:Correct_region} over this set gives \eqref{eq:Correct_region_weighting} to reduce CVaR:
\begin{equation}
	\label{eq:Correct_region_weighting}
	\setlength\abovedisplayskip{1pt}
	\setlength\belowdisplayskip{2pt}
	\begin{aligned}
		&L_{\alpha\%,t}^\mathrm{CVaR} - L_{t}^\mathrm{EME} \le &\frac{\sum\limits_{s \in \Omega_{\mathrm{REP},t}^{\mathrm{CVaR}}}
			\sum\limits_{e \in \Omega _\mathrm{SES}} G_{e,s}\left[\hat E_{e,t}^\mathrm{SES}-E_{e,t}^\mathrm{SES}\right]}{\lceil (1-\alpha\%)\cdot N_\mathrm{REP}\rceil}
	\end{aligned}
\end{equation}

The \eqref{eq:Correct_region_weighting} provides a linearized condition for reducing CVaR through iterative updates of the long-term SoC sequence. This facilitates a feedback-driven mitigation loop, enabling tail risk mitigation to long-term dispatch strategies.

\section{Case Studies}
\subsection{Set-Up}
Case studies are conducted on a modified IEEE-39 bus system featuring high RE penetration. The system retains seven TPSs located at buses 31–37, eight RE stations, and a single SES unit. The RE stations include 4 wind farms and 4 PV stations, each rated at 600 MW. The main parameters of the test system are summarized in Table \ref{Tab:Case_parameters}, other parameters follow the original IEEE-39 system specifications.

Historical records for renewable generation are sourced from publicly available data of Berlin, Bremerhaven, Dresden, and Düsseldorf, covering 30 years from 1980 to 2009.  These datasets construct the historical scenario set $\Omega_\mathrm{HST}$ for the wind and PV stations, with $\left|\Omega_\mathrm{T}\right|=8760$ hours. 
Likewise, load demand curves reflecting relatively high electricity demand conditions are developed using historical German load records from the same years. These data are available via \cite{cadillack_renewable_2025}. The parameters of Algorithm \ref{alg:scenario_generation} are set as: $N_\mathrm{REP}=200$, $A_\mathrm{G}=365$ hours, and $K_\mathrm{G}=24$. Additionally, the weight coefficient $\omega^\mathrm{RT}$ in \eqref{eq:Evolution_obj} is set to 10, and the CVaR risk threshold $T_{\alpha\%,t}^\mathrm{CVaR}$ is specified as $2\times10^5$\$.

The case studies are coded in MATLAB using the YALMIP interface and solved with the GUROBI 12.0 solver. The programming environment is Core i7-12700H @ 2.70GHz with 16 GB RAM.

\begin{table}[htb]
	\footnotesize
	\caption{Main parameters of the system.}
	\label{Tab:Case_parameters}
	\setlength{\tabcolsep}{3pt}
	\centering
	\renewcommand\arraystretch{0.9} 
	\begin{tabular}{p{45pt}<{\centering} m{65pt}<{\centering} m{45pt}<{\centering} m{65pt}<{\centering}}
		\toprule
		Parameters &
		Value&	
		Parameters & 
		Value
		\\[0pt]
		\midrule
		$\overline{P}_{e}^{\mathrm{SES}}$&	1000MW &	$\overline{E}_{e}^{\mathrm{SES}}$&	80000MWh
		\\[3pt]
		$w_e^\mathrm{D}$ &0.05 &$w_e^\mathrm{U}$ &0.95	  
		\\[3pt]
		$\eta_e^\mathrm{CH}$&	70\% &$\eta_e^\mathrm{DC}$&	70\%
		\\[3pt]
		$E_{\mathrm{INT},e}^\mathrm{SES}$ &	40000MWh &$\overline{P}_{m}^\mathrm{DR}$ & 2\% of the total LD
		\\[3pt]
		$\alpha$\% & 80\% &$c_{e}^{{\rm{SES}}}$ &51\$/MWh	
		\\[3pt]
		$c_{m}^{{\rm{DR}}}$ &100\$/MWh  &	$c^\mathrm{EME}$  &	1000\$/MWh
		\\[3pt]
		$\Delta t$ & 1 hour & $\delta^\mathrm{R}$& 4 hours
		\\[0pt]
		\bottomrule
	\end{tabular}
\end{table}

\begin{figure}[htb]
	\centering
	\includegraphics[width=8.6cm]{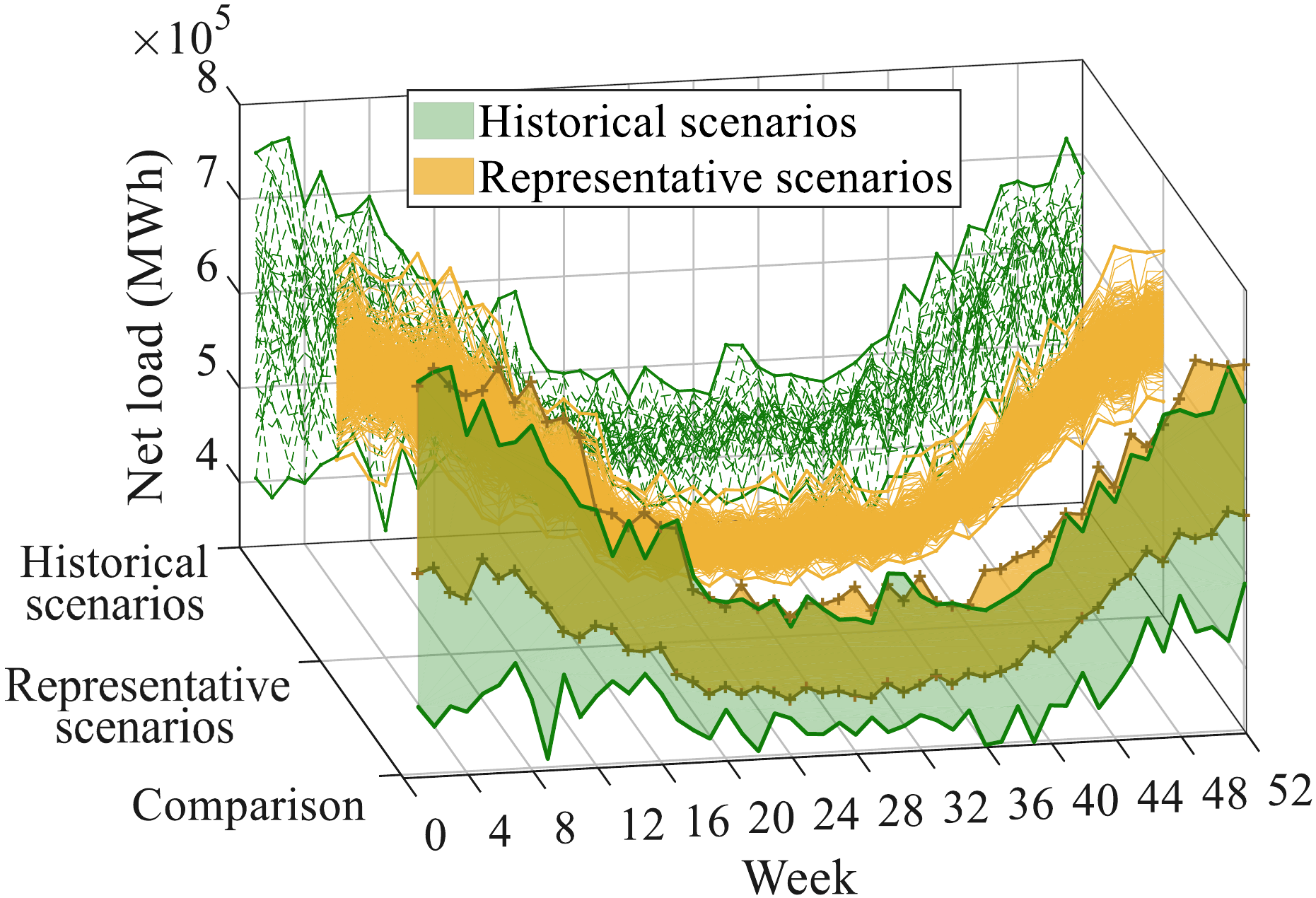}
	\caption{Comparison of the net load between historical scenario set $\Omega_\mathrm{HST}$ and representative scenario set $\Omega_\mathrm{REP}$.}
	\label{Fig:Load_renewable_curve}
\end{figure}

\begin{figure}[htb]
	\centering
	\subfloat[SoC sequence of SES]{\label{Fig:SoC_and_CVaR:a}
		\includegraphics[width=8.95cm]{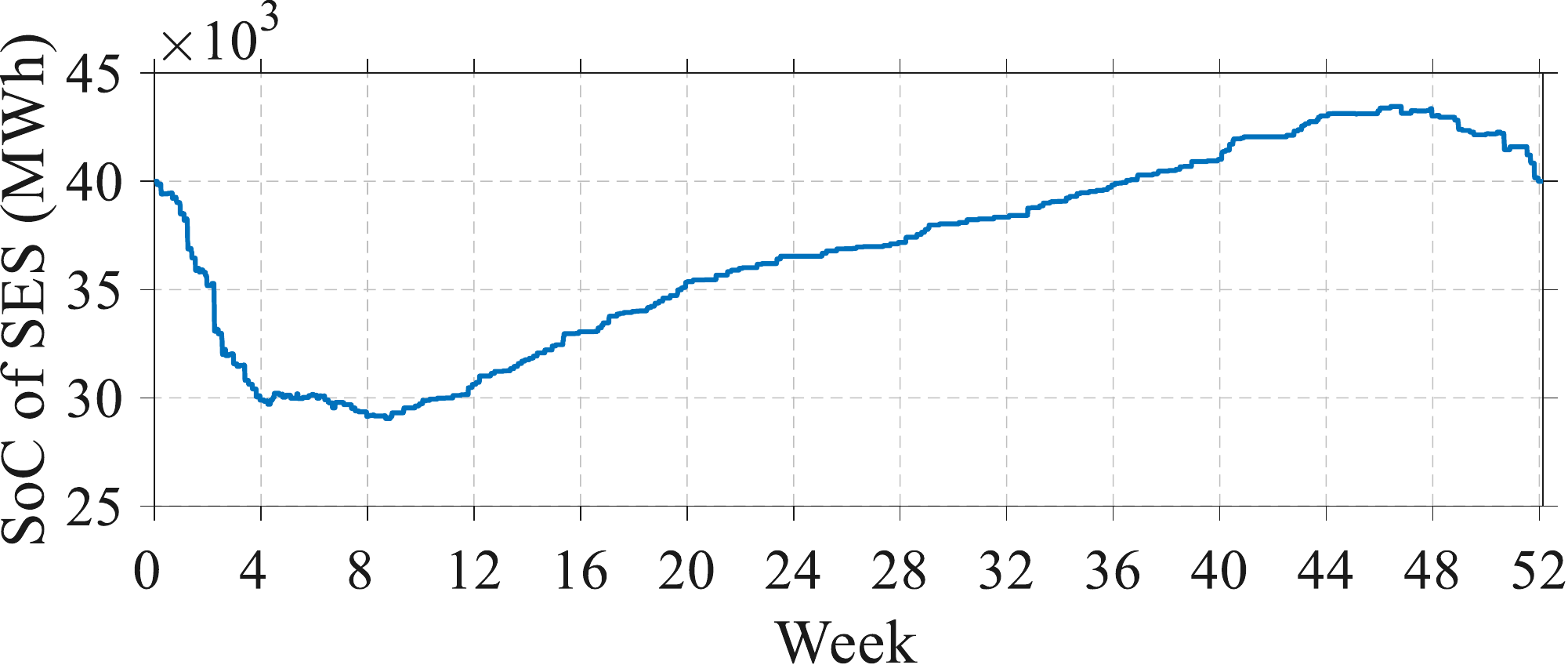}}\\ 
	\subfloat[CVaR of the emergency supply cost]{\label{Fig:SoC_and_CVaR:b}
		\includegraphics[width=8.95cm]{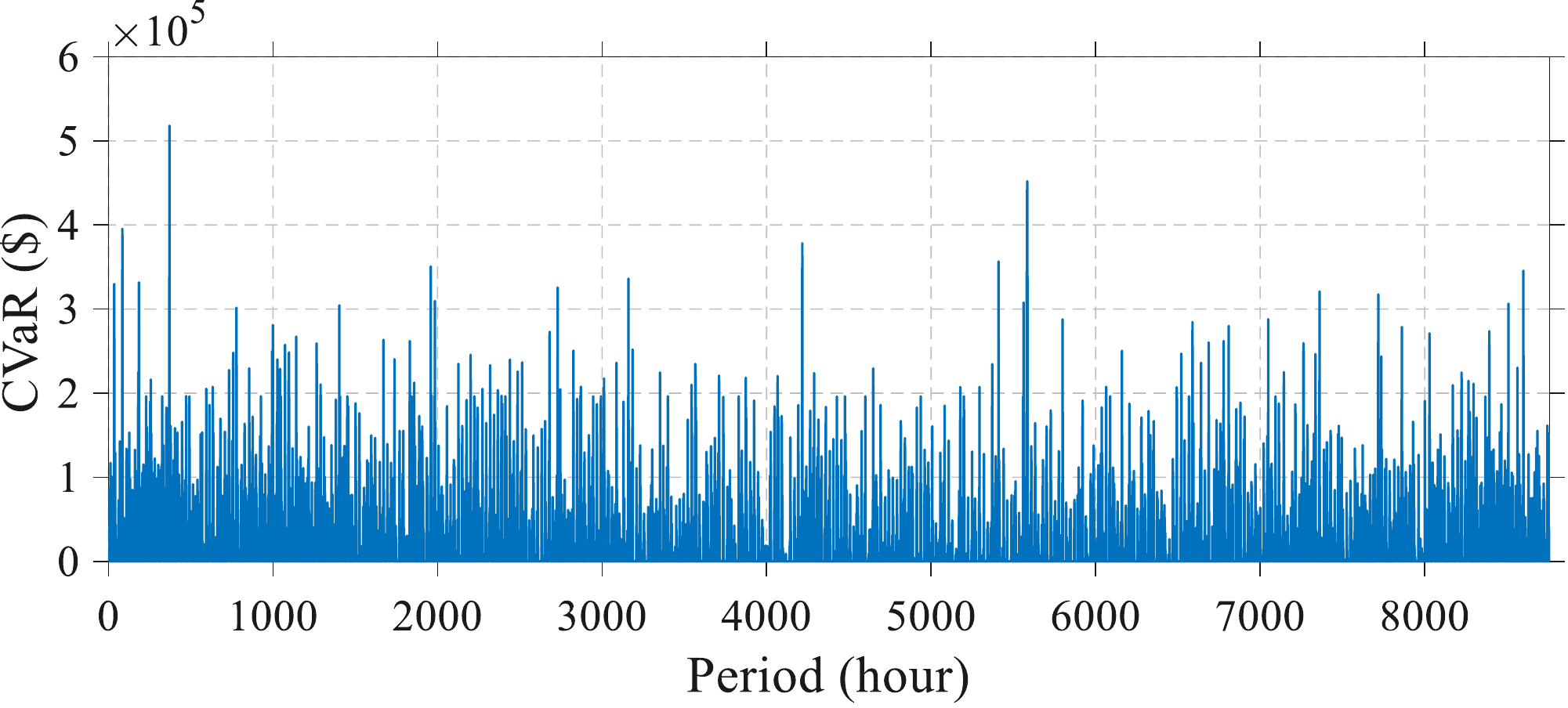}}\\
	\subfloat[Subgradient when $L_{\alpha\%,t}^\mathrm{CVaR} > T_{\alpha\%,t}^\mathrm{CVaR}$]{\label{Fig:SoC_and_CVaR:c}
			\includegraphics[width=8.95cm]{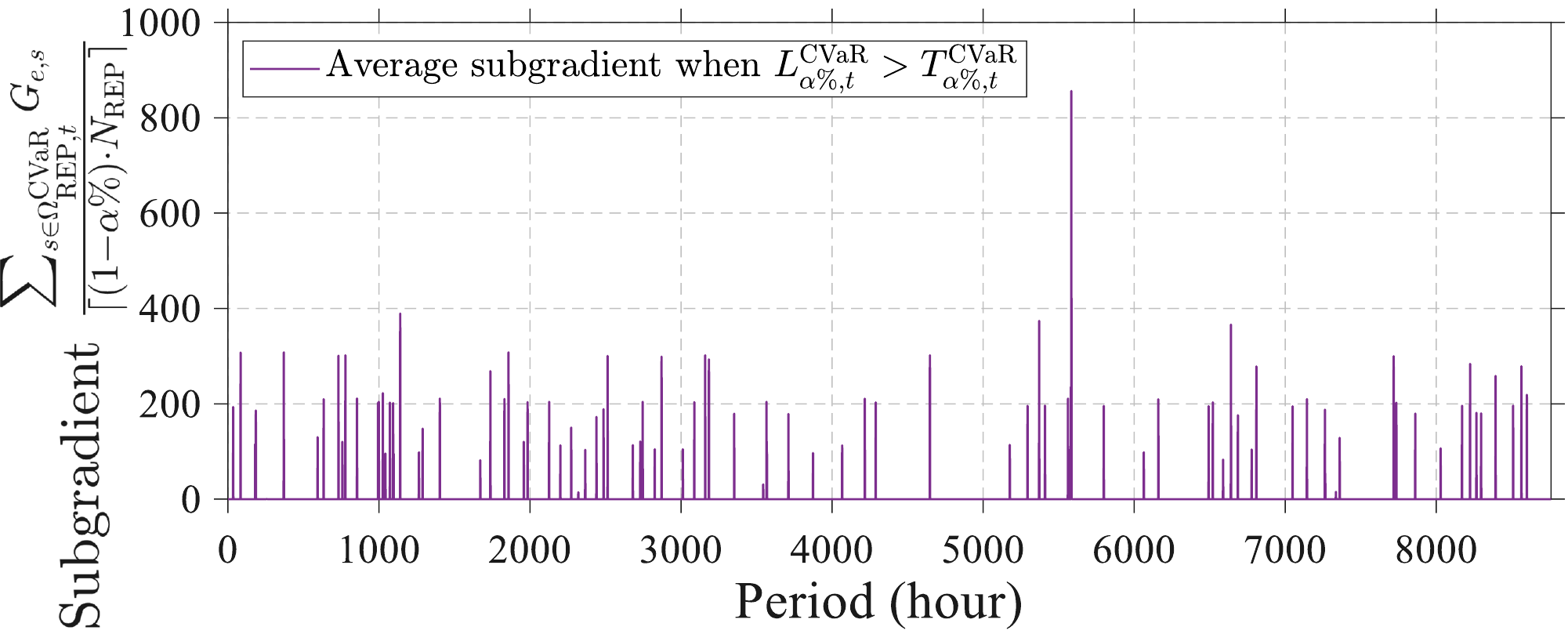}}
	\caption{SoC sequence, CVaR, and subgradient of the test case.}
	\label{Fig:SoC_and_CVaR} 
\end{figure}

\subsection{Representative Scenario Generation}
Fig. \ref{Fig:Load_renewable_curve} compares the net load between the historical scenario set and the representative scenario set generated by Algorithm \ref{alg:scenario_generation}.  By leveraging the multi-timescale Copula-based scenario generation approach, the representative scenarios capture coupled, potentially persistent power shortages in renewable power systems. Notably, although some extreme conditions have not been observed historically, the proposed method proactively includes these plausible risks in the assessment framework.

\subsection{Risk Assessment and Mitigation Analysis}
Fig. \ref{Fig:SoC_and_CVaR} illustrates the SoC sequence of the SES, the results of the CVaR-based long-term risk assessment, and the average subgradient values associated with periods when the CVaR $L_{\alpha\%,t}^{\mathrm{CVaR}}$ exceeds the predefined risk threshold $T_{\alpha\%,t}^{\mathrm{CVaR}}$. 
Fig. \ref{Fig:SoC_and_CVaR:a} displays the long-term dispatch strategy for the SES. This strategy, combined with the short-term decision-making model \eqref{eq:Evolution}, serves as the basis for the risk evaluation.  Fig. \ref{Fig:SoC_and_CVaR:b} demonstrates that under the current dispatch strategies, several periods experience elevated risk, indicated by instances where $L_{\alpha\%,t}^{\mathrm{CVaR}} > T_{\alpha\%,t}^{\mathrm{CVaR}}$. 
Importantly, the average subgradient computed according to Section \ref{Section:Risk_Mitigation} satisfies $\frac{\sum\nolimits_{s \in \Omega_{\mathrm{REP},t}^{\mathrm{CVaR}}}G_{e,s}}{\lceil (1-\alpha\%)\cdot N_\mathrm{REP}\rceil}>0$ under the above risk events.
 Correspondingly, constraint (12) suggests increasing the SES discharge level during these high-risk periods, thereby enhancing the system's ability to cope with risk events.

\section{Conclusions}
This paper presents a framework for long-term tail risk assessment and mitigation in renewable power systems, addressing the challenges posed by the seasonality and extremity of RE, as well as the complexity of incorporating long-term dispatch strategies. 
A multi-timescale Copula-based scenario generation method captures the long-range correlation and variability of RE and load demand. Building upon these scenarios, an evolution-based risk assessment model is developed, which integrates CVaR to quantify tail risks. A controlled evolution-based mitigation strategy is then introduced to refine long-term dispatch strategies and mitigate tail risks. Case studies on a modified IEEE-39 bus system demonstrate the effectiveness of the proposed framework in accurately evaluating tail risks and achieving risk mitigation through improved dispatch strategies. Although the proposed framework shows promising effectiveness, this study remains preliminary. Future research will focus on improving the accuracy of representative scenario generation under forecast information, reducing the computational complexity of long-term risk assessments, and exploring more efficient feedback schemes for enhancing complex dispatch strategies.

\section*{Acknowledgment}
This work was supported by the Joint Research Fund in Smart Grid (No.U1966601) under cooperative agreement between the National Natural Science Foundation of China (NSFC).



\ifCLASSOPTIONcaptionsoff
\newpage
\fi
\bibliographystyle{IEEEtran}
\bibliography{IEEEabrv,mybib}

\end{document}